\documentclass[amsmath, amssymb, aps, prd, nofootinbib, onecolumn, superscriptaddress
]{revtex4-2}

\usepackage{bm}
\usepackage{color}
\usepackage{dcolumn}
\usepackage{empheq}
\usepackage{graphicx}
\usepackage{hyperref}
\usepackage[normalem]{ulem}
\usepackage[dvipsnames]{xcolor}

\usepackage{tipa} 
\newcommand{\thorn}{\mbox{\textthorn}} 

\hypersetup{
    colorlinks = true,
    citecolor  = blue,
    linkcolor  = red,
}

\newcommand{\be}{\begin{equation}}
\newcommand{\ee}{\end{equation}}
\newcommand{\bear}{\begin{eqnarray}}
\newcommand{\eear}{\end{eqnarray}}

\newcommand\blfootnote[1]{%
  \begingroup
  \renewcommand\thefootnote{}\footnote{#1}%
  \addtocounter{footnote}{-1}%
  \endgroup
}

\begin{document}

\title{Penrose--Rindler equation and horizon thermodynamics of stationary black holes}

\author{Diego Fernández-Silvestre}
\email{dfsilvestre@ubu.es}
\affiliation{Departamento de Matemáticas y Computación, Universidad de Burgos, 09001 Burgos, Spain}

\author{Alberto Guilabert}
\email{alberto.guilabert@ua.es}
\affiliation{Departamento de Física, Universidad de Alicante, Campus de San Vicente del Raspeig, E-03690 Alicante, Spain}

\author{Pedro Bargueño}
\email{pedro.bargueno@ua.es}
\affiliation{Departamento de Física, Universidad de Alicante, Campus de San Vicente del Raspeig, E-03690 Alicante, Spain}

\author{Juan A. Miralles}
\email{ja.miralles@ua.es}
\affiliation{Departamento de Física, Universidad de Alicante, Campus de San Vicente del Raspeig, E-03690 Alicante, Spain}

\date{\today}

\begin{abstract}
    Black holes are the natural arena for exploring the interplay between gravity and thermodynamics. Although the association between black hole mechanics and black hole thermodynamics is well established, the comprehensive geometric formulation of thermodynamic variables deserves further investigation. In this work, both Newman--Penrose (NP) and Geroch--Held--Penrose (GHP) formalisms are considered within the framework of horizon thermodynamics. We show that the NP formalism reformulates the horizon condition as the Penrose--Rindler equation. In this context, a Smarr-like formula for stationary black holes is recovered from the Penrose--Rindler equation reinterpreted as a horizon equilibrium of pressures, which includes a pressure associated with the horizon rotation. A complete geometric reformulation of this reinterpretation of the Penrose--Rindler equation evaluated at the horizon is developed within the GHP formalism. The GHP approach further inspires the introduction of the horizon-averaged matter pressure and its conjugate volume, thereby enabling a quasilocal realization of the Smarr-like formula for stationary black holes. This geometric formulation clarifies the connection between horizon dynamics and thermodynamics and offers a unified setting for extending black hole thermodynamics beyond spherical symmetry.
\end{abstract}


\blfootnote{\bigskip}

\maketitle

\section{Introduction}

Black holes, apart from being exact solutions of Einstein’s field equations, can be consistently interpreted as thermodynamic systems. The black hole area theorem, which ensures the existence of an irreducible mass \cite{Christodoulou:1970wf}, bears a close resemblance to the concept of entropy in conventional thermodynamics. The black hole area was definitely associated with its entropy by Bekenstein \cite{Bekenstein:1973ur}. All these developments allowed the laws of black hole mechanics to be reinterpreted as the laws of black hole thermodynamics \cite{Bardeen:1973gs}, with the first law yielding the Smarr formula upon integration \cite{Smarr:1972kt}. This analogy became concrete with the discovery of Hawking radiation, which established that black holes indeed possess temperature and entropy \cite{Hawking:1975vcx}.

Black hole thermodynamics was traditionally investigated using global properties of event horizons in asymptotically flat spacetimes. Its global character, however, makes the notion of the event horizon not well defined in some cases, particularly in cosmological or dynamical spacetimes. This limitation motivated the introduction of quasilocal approaches, and the concept of trapping horizons provided a more accessible and a general definition of black holes based on local geometric conditions \cite{Hayward1994}.

According to the apparent thermodynamic behavior of black holes, many efforts have been dedicated to describing them as genuine thermodynamic systems. As a summary, there are two main approaches to the study of black hole thermodynamics: horizon and extended phase space. The horizon thermodynamics approach is based on the local interpretation of the radial Einstein's field equation evaluated at the horizon as a thermodynamic identity related to the first law of black hole thermodynamics. The pressure is associated with the radial component of the energy-momentum tensor (including a possible cosmological constant), and the volume is the conjugate variable to the matter pressure in a formal or geometric sense \cite{Jacobson:1995ab, Hayward:1997jp, Padmanabhan:2002sha, Padmanabhan:2003gd, Kothawala:2007em, Padmanabhan:2009vy, Padmanabhan:2009kr, Hansen:2016ayo, Hansen:2016gud, Hansen:2016wdg}. The extended phase space approach is about the global interpretation of black holes as thermodynamic systems with the cosmological constant treated as a thermodynamic variable. The pressure is associated with the cosmological constant, and the so-called thermodynamic volume is the conjugate variable to the cosmological pressure in the standard thermodynamic sense \cite{Kastor:2009wy, Dolan:2010ha, Kubiznak:2012wp, Dolan:2012jh, Dolan:2014jva, Kubiznak:2016qmn, Mann:2025xrb}. In both contexts, the idea of assigning an equation of state to black holes has been particularly successful, enabling, for example, the construction of phase diagrams and the study of phase transitions as typically done in standard thermodynamic systems. Additionally, black hole equations of state may be further interpreted in terms of ``spacetime atoms'' or horizon degrees of freedom, which might be helpful in investigating the analogy between gravity and thermodynamics using statistical mechanics techniques, possibly shedding light on the microscopic structure underlying the Bekenstein--Hawking entropy \cite{Wei:2015iwa, Vargas:2017eta, Villalba:2020acp}. 

This manuscript lies within the framework of horizon thermodynamics. A central challenge in this program is to go deep into the thermodynamic counterpart of purely geometric quantities defined at the horizon. In this context, quasilocal geometry is capable of capturing geometric information intrinsic to horizons, potentially rendering a new perspective on the horizon approach to black hole thermodynamics.

An appropriate and convenient framework for developing these ideas is provided by the Newman--Penrose (NP) formalism or the more recent Geroch--Held--Penrose (GHP) formalism \cite{Newman:1961qr, Geroch:1973am, Penrose1984, Chandrasekhar1998, Bargueno2023}. In these formalisms, the Riemann curvature tensor is expressed in terms of scalar invariants and spin coefficients associated with a null tetrad. Notably, for algebraically special spacetimes of Petrov type D, the two real null vectors in the tetrad basis can be chosen as adapted to the principal null directions of the Weyl tensor. This framework thus provides natural candidates for describing the geometry of horizon null surfaces governing the thermodynamics of black holes. A central equation of these formalisms is the Penrose--Rindler equation \cite{Penrose1984} (see also \cite{Bargueno2023, Villalba:2020dbv, Guilabert:2024tga}), which will play a relevant role in this work. It will allow a physical reinterpretation of black hole horizons as equilibria of competing pressures in a way reminiscent of interacting fluids. Finally, the connection between geometry and thermodynamics will be elucidated by relating the Penrose--Rindler equation with the generalized version of the Smarr formula \cite{Smarr:1972kt}.

This manuscript is organized as follows: In Section \ref{Sec2}, we analyze static and spherically symmetric black holes within the NP formalism, demonstrating that the horizon condition is reformulated as the Penrose--Rindler equation and reinterpreted as an equilibrium between matter, curvature, and thermal pressures. A van der Waals-like equation of state naturally emerges through the holographic energy equipartition of the horizon. The Penrose--Rindler equation is finally related to the Smarr formula. In Section \ref{Sec3}, we extend this framework to stationary and axisymmetric spacetimes, where the Weyl NP scalar becomes complex. The Penrose--Rindler equation is again shown to coincide with the horizon condition, and a pressure related to rotation enters the horizon equilibrium. The Penrose--Rindler equation is also related to a generalized Smarr formula for Kerr-like geometries that includes the angular momentum and the angular velocity of the horizon. In Section \ref{Sec4}, we reformulate the analysis within the GHP formalism, which provides a generalized and geometrically transparent decomposition of the pressure into curvature, thermal, and rotational terms. The GHP formalism motivates, inspired by quasilocal geometry, the introduction of the Smarr volume, a horizon-averaged quantity that acts as the conjugate of the horizon-averaged matter pressure, thus recovering a Smarr-like relation for stationary black holes from geometric considerations. Finally, Section \ref{Sec5} summarizes this work. Throughout this manuscript, we adopt the metric signature $(1, 3)$ and the Penrose--Rindler sign convention for the Riemann curvature tensor \cite{Penrose1984}.

\section{Thermodynamics of static black holes within the NP formalism} \label{Sec2}

We begin by analyzing static and spherically symmetric black holes within the NP formalism. We focus, in particular, on spacetimes with the following ansatz for the line element: 
\begin{equation} \label{SSMetric}
    ds^2 = f(r) dt^2 - f(r)^{-1} dr^2 - r^2 (d\theta^2 + \sin^2 \theta d\phi^2),
\end{equation}
which is completely specified in terms of the function $f(r)$. Note that this case does not describe the most general static spacetime with spherical symmetry, but many relevant solutions are of this type, remarkably, the Schwarzschild and Reissner--Nordström spacetimes.

As commented in the Introduction, the concept of a black hole horizon is subtle, since the event horizon is defined globally in terms of the causal structure of the spacetime. Building on this, Hayward proposed a quasilocal characterization of black holes in terms of future outer trapping horizons, which encompasses the concept of dynamical horizon even in nonstationary settings \cite{Hayward1994}. In order to analyze the geometric and thermodynamic properties of trapping horizons, the NP formalism provides a natural framework for expressing geometric notions, such as expansion, rotation, shear, and surface gravity, in terms of spin coefficients \cite{Newman:1961qr, Geroch:1973am, Penrose1984, Chandrasekhar1998, Bargueno2023}. As the considered spacetime is of Petrov type D, we define the null tetrad with the two real null vectors adapted to the two principal null directions of the Weyl tensor. The null tetrad is then uniquely determined, up to proper scaling factor. It is given by
\begin{equation}
    \begin{aligned}
        l^\mu &= \frac{1}{\sqrt{2}} \left(f(r)^{-1}, 1, 0, 0\right), \\
        n^\mu &= \frac{1}{\sqrt{2}} \left(1, - f(r), 0, 0\right), \\
        m^\mu &= \frac{1}{\sqrt{2} r} \left(0, 0, 1, i \csc\theta \right),
    \end{aligned}
\end{equation}
where $l_\mu n^\mu = 1$ and $m_\mu \overline{m}^\mu = - 1$, with the bar denoting complex conjugation. The null tetrad is such that the null vectors $\mathbf{l}$ and $\mathbf{n}$ are then aligned with the outgoing and ingoing null directions, respectively.

In this tetrad basis, the NP scalars and spin coefficients can be computed straightforwardly. The only nonvanishing NP scalars are
\begin{equation} \label{NP}
    \begin{aligned}
        \Psi_2 &= C_{\mu \nu \rho \sigma} l^\mu m^\nu \overline{m}^\rho n^\sigma, \\
        \Phi_{11} &= - \frac{1}{2} R_{\mu \nu} l^\mu n^\nu + 3 \Lambda, \\
        \Lambda &= \frac{R}{24}, \\
    \end{aligned}
\end{equation}
where $C_{\mu \nu \rho \sigma}$ and $R_{\mu \nu}$ stand for the components of the Weyl and Ricci tensors, and $R = g^{\mu \nu} R_{\mu \nu}$ denotes the Ricci scalar.

Note that the three nonvanishing NP scalars will depend only on the metric function $f(r)$, and its first and second derivatives $f'(r)$ and $f''(r)$. As a consequence, it is possible to solve the system of equations and rewrite the metric function and its derivatives in terms of the NP scalars as
\begin{align}
    f(r) &= 1 + 2 r^2 \left(\Psi_2 - \Phi_{11} - \Lambda\right), \label{SSNP1} \\
    f'(r) &= -  2 r \left(\Psi_2 + 2 \Lambda\right), \label{SSNP2} \\
    f''(r) &= 4 \left(\Psi_2 + \Phi_{11} - \Lambda\right). \label{SSNP3}
\end{align}

The black hole horizon condition is $f(r) \,\hat{=}\, 0$,\footnote{The symbol $\hat{=}$ denotes equality at the horizon, \textit{i.e.}, evaluated at the horizon null surface $r = r_H$.} so Eq.~\eqref{SSNP1} gives, in terms of the NP scalars,
\begin{equation}
    - \Psi_2 + \Phi_{11} + \Lambda \,\hat{=}\, \frac{1}{2 r^2},
\end{equation}
which, interestingly, corresponds to the Penrose--Rindler equation \cite{Penrose1984} for static and spherically symmetric spacetimes evaluated at the horizon, usually written as
\begin{equation} \label{PR}
    - \Psi_2 + \Phi_{11} + \Lambda \,\hat{=}\, \frac{k_g}{2},
\end{equation}
where $k_g$ denotes the Gaussian curvature of the orthogonal surface to $\mathbf{l}$ and $\mathbf{n}$. In this case, the black hole horizon can be foliated by marginally trapped two-spheres, thus,
\begin{equation}
    k_g \,\hat{=}\, \frac{1}{r^2}.
\end{equation}
We recall that the Gaussian curvature can also be expressed in terms of the so-called Penrose--Rindler (complex) $K$-curvature \cite{Penrose1984} as
\begin{equation} \label{K}
    k_g = K + \overline{K}.
\end{equation}

The black hole temperature is $T \,\hat{=}\, \frac{f'(r)}{4 \pi}$, so Eq.~\eqref{SSNP2} gives, in terms of the NP scalars,
\begin{equation} \label{SST}
    T \,\hat{=}\, - \frac{1}{2 \pi} r \left(\Psi_2 + 2 \Lambda\right).
\end{equation}

Let us now write the Penrose--Rindler equation \eqref{PR} as
\begin{equation} \label{PR2}
    - (\Psi_2 + 2 \Lambda) + \Phi_{11} + 3 \Lambda \,\hat{=}\, \frac{k_g}{2}.
\end{equation}
This leads to a more physical and simple interpretation of static black holes with spherical symmetry as defined by an equilibrium of horizon pressures given by
\begin{equation} \label{SSPEq}
    P \,\hat{=}\, P_T + P_{k_g},
\end{equation}
where $P = - T_r^r$ is the pressure associated with the radial component of the energy-momentum tensor (including a possible cosmological constant), that is, the matter pressure, which reads\footnote{The $4 \pi$ factor arises from the definition of the energy-momentum tensor. The minus sign is a consequence of the conventions for the metric signature and the Riemann tensor.}
\begin{equation} \label{P}
    P = - \frac{1}{4 \pi} (\Phi_{11} + 3 \Lambda).
\end{equation}
We have also introduced $P_T$, which is related to the temperature in Eq.~\eqref{SST}. We will call it thermal pressure and it reads
\begin{equation} \label{PT}
    P_T \,\hat{=}\, - \frac{1}{4 \pi} (\Psi_2 + 2 \Lambda).
\end{equation}
Finally, $P_{k_g}$ is the pressure associated with the Gaussian curvature, the so-called curvature pressure \cite{Hansen:2016ayo}, which reads
\begin{equation} \label{PKG}
    P_{k_g} = - \frac{k_g}{8 \pi}.
\end{equation}

In fact, it is possible to give a microscopic interpretation of these pressures by introducing the holographic energy equipartition, more specifically by invoking the holographic degrees of freedom $N$ together with the areal or geometric volume of the horizon $V$ (the volume of the two-sphere). This volume coincides with the coined as thermodynamic volume defined for black holes with spherical symmetry in the extended phase space thermodynamics \cite{Kubiznak:2016qmn}. By the holographic equipartition principle $N = A / \ell_p^2$, the Penrose--Rindler equation \eqref{PR} can be recast as a black hole equation of state $P \,\hat{=}\, P(n, T)$, with $n = \bar{N} / V$ and $\bar{N} = N / 6$. Accordingly, the equation of state takes the form
\begin{equation} \label{EoS}
    P(n, T) \,\hat{=}\, n T - \frac{n^2}{2 \pi}.
\end{equation}

While Eq.~\eqref{SSPEq} allows for the fundamental physical interpretation of the horizon condition as an equilibrium of pressures, 
Eq.~\eqref{EoS} should be understood as the appropriate reformulation in terms of the holographic degrees of freedom of the horizon, not as an independent physical interpretation. A static and spherically symmetric black hole can be consequently interpreted in terms of a van der Waals-like equation of state, such that the thermal pressure can be associated with the pressure corresponding to the kinetic energy of an ideal gas and the curvature pressure with the pressure corresponding to a van der Waals interaction term.

At this point, we have shown that the black hole horizon condition is geometrically equivalent to implying the Penrose--Rindler equation, which is just the Penrose and Rindler theorem for the (complex) $K$-curvature evaluated at the horizon. This was already shown slightly differently in \cite{Villalba:2020dbv}. Additionally, the Penrose--Rindler equation can be physically interpreted in terms of an equilibrium of pressures at the horizon. In this context of horizon thermodynamics and once the holographic energy equipartition of the horizon is introduced, the Penrose--Rindler equation has shown to reproduce a van der Waals-like equation of state for spherically symmetric and static black holes. This was already obtained in \cite{Vargas:2017eta, Villalba:2020dbv} for the Schwarzschild and Reissner--Nordström black holes contained in an anti-de Sitter spacetime. We have arrived at exactly the same result, but without resorting to a concrete solution. This also holds, for instance, for black hole solutions in de Sitter spacetime when restricting to the thermodynamics of a single horizon. One usually considers the black hole horizon, but the same should work for the cosmological horizon. In the case of Schwarzschild-anti-de Sitter spacetime specifically, the holographic degrees of freedom were interestingly realized through a microscopic interaction model that reproduced the Bekenstein--Hawking entropy and the van der Waals equation of state using statistical mechanics techniques \cite{Vargas:2017eta}.

What is more, the Penrose--Rindler equation is equivalent to the Smarr formula in the static and spherically symmetric case. One can check this by multiplying Eq.~\eqref{PR2} by $3 V$ and noting that $3 P_T V \,\hat{=}\, 2 T S $ and $3 P_{k_g} V \,\hat{=}\, - E$, hence,
\begin{equation} \label{SSS}
    E \,\hat{=}\, 2 T S - 3 P V,
\end{equation}
where $V \,\hat{=}\, \frac{4 \pi}{3} r^3$ denotes the areal or geometric volume of the black hole. The above equation is not exactly the celebrated Smarr formula \cite{Smarr:1972kt}, but it generalizes it for arbitrary static black holes, where
\begin{equation}
    E \,\hat{=}\, \frac{r}{2}, \quad S = \frac{A}{4} \,\hat{=}\, \pi r^2
\end{equation}
are the horizon energy (Misner--Sharp energy) and entropy, respectively. The reader is referred to \cite{Guilabert:2024tga} for a more sophisticated treatment using the GHP formalism. This connection between the Penrose--Rindler equation and the Smarr formula will be extended to stationary and axisymmetric spacetimes in the following sections.

\section{Thermodynamics of stationary black holes within the NP formalism} \label{Sec3}

Let us consider, for simplicity, stationary and axisymmetric rotating (Kerr-like) spacetimes with the following ansatz for the metric:
\begin{equation} \label{ASMetric}
    ds^2 = \frac{\Delta(r)}{\rho^2} (dt - a \sin^2 \theta d\phi)^2 - \frac{\rho^2}{\Delta(r)} dr^2 - \rho^2 d\theta^2 - \frac{\sin^2 \theta}{\rho^2} (a dt - (r^2 + a^2) d\phi)^2,
\end{equation}
in terms of the function $\Delta(r)$, and where $\rho^2 = r^2 + a^2 \cos^2 \theta$ with $a$ the rotation parameter. Again, this metric is not the most general one regarding axial symmetry and rotation but encompasses, importantly, the Kerr family of asymptotically flat black hole solutions.\footnote{Asymptotically (anti-)de Sitter solutions are not described by this metric, but the results generalize straightforwardly.} This ansatz has already been considered in the literature, in particular, in the context of horizon thermodynamics \cite{Kothawala:2007em, Hansen:2016wdg}. Notice that the ansatz for static spacetimes with spherical symmetry \eqref{SSMetric} follows from the limit $a \to 0$ provided the identification $\Delta(r) = r^2 f(r)$.

Again, as the considered spacetime is of Petrov type D, we define the null tetrad with the two real null vectors adapted to the two principal null directions of the Weyl tensor. It is given by
\begin{equation}
    \begin{aligned}
        l^\mu &= \frac{1}{\sqrt{2} \Delta(r)} \left(r^2 + a^2, \Delta(r), 0, a\right), \\
        n^\mu &= \frac{1}{\sqrt{2} \rho^2} \left(r^2 + a^2, - \Delta(r), 0, a\right), \\
        m^\mu &= \frac{1}{\sqrt{2} ( r + i a \cos\theta)} \left(i a \sin\theta, 0, 1, i \csc\theta \right),
    \end{aligned}
\end{equation}
where $l_\mu n^\mu = 1$ and $m_\mu \overline{m}^\mu = - 1$. As before, the null tetrad is such that the null vectors $\mathbf{l}$ and $\mathbf{n}$ are then aligned with the outgoing and ingoing null directions, respectively.

In this case, the nonvanishing NP scalars are also those in Eq.~\eqref{NP}. However, unlike the static case where the Weyl scalar $\Psi_2$ is real (as well as the Ricci scalars $\Phi_{11}$ and $\Lambda$), in the stationary case the Weyl scalar $\Psi_2$ is complex. The imaginary part of $\Psi_2$ is usually associated with rotation, since it vanishes in the limit $a \to 0$. As before, the three nontrivial NP scalars will depend only on the metric function $\Delta(r)$ and its first and second derivatives. We can therefore solve this system of equations and obtain $\Delta(r)$ and its derivatives in terms of the NP scalars as
\begin{equation}
    \begin{aligned}
        \Delta(r) &= 2 r^4 (\Psi_2 - \Phi_{11} - \Lambda) - 2 i r^3 a \cos\theta (3 \Psi_2 - 2 \Phi_{11}) + r^2 (1 - 3 a^2 (1 + \cos2\theta) (\Psi_2 + 2 \Lambda) ) \\
        &\quad + 2 i r a^3 \cos^3\theta  (\Psi_2 + 2 \Phi_{11} + 8 \Lambda) + a^2 (1 + \frac{a^2}{4} (3 + 4 \cos2\theta + \cos4\theta) ) (\Phi_{11} + 3 \Lambda), \\
        \Delta'(r) &= 2 (r + (r - i  a \cos\theta)^2 (r (\Psi_2 - 2 \Phi_{11} - 4 \Lambda) - i a \cos\theta (\Psi_2 + 2 \Phi_{11} + 8 \Lambda) ) ), \\
        \Delta''(r) &= 2 (1 - 12 \rho^2 \Lambda).
    \end{aligned}
\end{equation}
The above expressions are highly nontrivial compared with Eqs.~\eqref{SSNP1}--\eqref{SSNP3} for spherically symmetric and static spacetimes. However, Eqs.~\eqref{SSNP1}--\eqref{SSNP3} are recovered when substituting $\Delta(r) = r^2 f(r)$ and taking the limit $a \to 0$. As already remarked, the Weyl scalar $\Psi_2$ is complex, so we write it as $\Psi_2 = \mathrm{Re}(\Psi_2) + i \mathrm{Im}(\Psi_2)$, thus obtaining
\begin{equation} \label{ASNP}
    \begin{aligned}
        \Delta(r) &= 2 r^4 (\mathrm{Re}(\Psi_2) - \Phi_{11} - \Lambda) + 6 r^3 a \cos\theta \mathrm{Im}(\Psi_2) + r^2 (1 - 3 a^2 (1 + \cos2\theta) (\mathrm{Re}(\Psi_2) + 2 \Lambda) ) \\ 
        &\quad - 2 r a^3 \cos^3\theta \mathrm{Im}(\Psi_2) + a^2 (1 + \frac{a^2}{4} (3 + 4 \cos2\theta + \cos4\theta) ) (\Phi_{11} + 3 \Lambda) \\
        &\quad + 2 i r C, \\
        \Delta'(r) &= 2 (r^3 (\mathrm{Re}(\Psi_2) - 2 \Phi_{11} - 4 \Lambda) + 3 r^2 a \cos\theta \mathrm{Im}(\Psi_2) + r (1 - a^2 \cos^2\theta (3 \mathrm{Re}(\Psi_2) + 2 \Phi_{11} + 12 \Lambda) ) - a^3 \cos^3\theta \mathrm{Im}(\Psi_2)) \\
        &\quad + 2 i C,
    \end{aligned}
\end{equation}
where
\begin{equation}
    C := r^3 \mathrm{Im}(\Psi_2) - r^2 a \cos\theta (3 \mathrm{Re}(\Psi_2) - 2 \Phi_{11}) - 3 r a^2 \cos^2\theta \mathrm{Im}(\Psi_2) + a^3 \cos^3\theta (\mathrm{Re}(\Psi_2) + 2 \Phi_{11} + 8 \Lambda),
\end{equation}
and the expression for $\Delta''(r)$ remains unchanged.

At this point, it becomes convenient to highlight a couple of points. On the one hand, the metric function $\Delta(r)$ is, of course, a real function. As a consequence, once the Weyl scalar is divided into its real and imaginary parts, the imaginary part of $\Delta(r)$ and $\Delta'(r)$ as given in Eq.~\eqref{ASNP} must be identically zero. We end up then with an algebraic relation between the NP scalars that states that $C = 0$. On the other hand, apart from not being complex, the metric function $\Delta(r)$ is a function of only the radial coordinate $r$. It is not a function of the angular coordinate $\theta$ specifically. Although Eq.~\eqref{ASNP} depends explicitly on $\theta$, the NP scalars depend implicitly on it, thus making the whole expression for $\Delta(r)$ and its derivatives $\theta$-independent. Note that also $C$ will be independent of this coordinate. 

What is more, the algebraic relation $C = 0$ tells us that, for the spacetimes of the type \eqref{ASMetric}, the NP scalars are not independent when the real and imaginary parts of the Weyl scalar are regarded separately. One can choose at this point to solve it and write one of the NP scalars in terms of the others. We choose to do it with respect to the imaginary part of the Weyl scalar so that every equation is given in terms of the real part of the Weyl scalar and the remaining Ricci scalars. This choice is natural in order to compare with the previously considered static spacetimes with spherical symmetry, where the Weyl scalar had only a real part. We have, then,
\begin{equation} \label{C}
    \mathrm{Im}(\Psi_2) = \frac{a \cos\theta (r^2 (3 \mathrm{Re}(\Psi_2) - 2 \Phi_{11}) - a^2 \cos^2\theta (\mathrm{Re}(\Psi_2) + 2 \Phi_{11} + 8 \Lambda) )}{r (r^2 - 3 a^2 \cos^2\theta)},
\end{equation}
which is compatible with the fact that $\mathrm{Im}(\Psi_2) \to 0$ in the limit $a \to 0$. In particular, this relation is reminiscent of the fact that the real and imaginary parts of the Weyl scalar are dependent on each other in the spacetimes under study. We recall that the real part is physically associated with tidal forces and the imaginary part is physically associated with frame dragging. An analysis of the real and imaginary parts of the Weyl scalar in stationary black holes and very nice visualizations of these effects can be found in \cite{Zhang:2012jj}. One can check Eq.~\eqref{C} with a particular solution, such as the Kerr spacetime, to convince oneself of this relation.

Finally, taking into account Eq.~\eqref{C} in Eq.~\eqref{ASNP},
\begin{equation} \label{ASNP2}
    \begin{aligned}
        \Delta(r) &= r^2 + a^2 + \frac{2 \rho^6}{r^2 - 3 a^2 \cos^2\theta} (\mathrm{Re}(\Psi_2) - \Phi_{11} - \Lambda), \\
        \Delta'(r) &= 2 r (1 + r^2 (\mathrm{Re}(\Psi_2) - 2 \Phi_{11} - 4 \Lambda) - a^2 \cos^2\theta (3 \mathrm{Re}(\Psi_2) + 2 \Phi_{11} + 12 \Lambda) ) \\
        &\quad + \frac{2 a^2 \cos^2\theta}{r (r^2 - 3 a^2 \cos^2\theta)} (3 r^4 (3 \mathrm{Re}(\Psi_2) - 2 \Phi_{11}) - 2 r^2 a^2 \cos^2\theta (3 \mathrm{Re}(\Psi_2) + 2 \Phi_{11} + 12 \Lambda) \\
        &\quad
        + a^4 \cos^4\theta (\mathrm{Re}(\Psi_2) + 2 \Phi_{11} + 8 \Lambda)).
    \end{aligned}
\end{equation}
Again, it should be stressed that these equations are independent of the angular coordinate. One can convince oneself of this fact, for instance, with the Kerr solution. Once the NP scalars are made explicit, the metric function $\Delta(r)$ will become that of Kerr, which is, of course, not dependent on $\theta$. The same applies to the corresponding derivatives.

We analyze these equations in light of horizon thermodynamics now. Our analysis is then based on stationary and axisymmetric rotating (Kerr-like) black holes, thus complementing and generalizing the previous section on static and spherically symmetric black holes.

Let us consider the first line in Eq.~\eqref{ASNP2}:
\begin{equation}
    \Delta(r) = r^2 + a^2 + \frac{2 \rho^6}{r^2 - 3 a^2 \cos^2\theta} (\mathrm{Re}(\Psi_2) - \Phi_{11} - \Lambda).
\end{equation}
The black hole horizon is located at $\Delta(r) \,\hat{=}\, 0$, so
\begin{equation}
    - \mathrm{Re}(\Psi_2) + \Phi_{11} + \Lambda \,\hat{=}\, \frac{k_g}{2},
\end{equation}
which again corresponds to the Penrose--Rindler equation \eqref{PR} (with only the real part of the Weyl scalar entering into it), where the Gaussian curvature of the horizon indeed is \cite{Smarr:1973zz}
\begin{equation} \label{ASKG}
    k_g \,\hat{=}\, \frac{(r^2 + a^2) (r^2 - 3 a^2 \cos^2\theta)}{\rho^6}.
\end{equation}
We remind that the Gaussian curvature can also be given in terms of the Penrose--Rindler (complex) $K$-curvature \eqref{K}. Note that the Gaussian curvature of the two-sphere is recovered in the limit of no rotation $a \to 0$. Although the constituents entering the Penrose--Rindler equation depend explicitly on the angular coordinate $\theta$, the equation itself is valid for all values of $\theta$, consistent with the fact that the black hole horizon condition $\Delta(r) \,\hat{=}\, 0$ is $\theta$-independent.

We move on to the second line in Eq.~\eqref{ASNP2}:
\begin{equation}
    \begin{aligned}
        \Delta'(r) &= 2 r (1 + r^2 (\mathrm{Re}(\Psi_2) - 2 \Phi_{11} - 4 \Lambda) - a^2 \cos^2\theta (3 \mathrm{Re}(\Psi_2) + 2 \Phi_{11} + 12 \Lambda) ) \\
        &\quad + \frac{2 a^2 \cos^2\theta}{r (r^2 - 3 a^2 \cos^2\theta)} (3 r^4 (3 \mathrm{Re}(\Psi_2) - 2 \Phi_{11}) - 2 r^2 a^2 \cos^2\theta (3 \mathrm{Re}(\Psi_2) + 2 \Phi_{11} + 12 \Lambda)\\
        &\quad
        + a^4 \cos^4\theta (\mathrm{Re}(\Psi_2) + 2 \Phi_{11} + 8 \Lambda) ).
    \end{aligned}
\end{equation}
The black hole temperature is $T \,\hat{=}\, \frac{\Delta'(r)}{4 \pi (r^2 + a^2)}$, so
\begin{equation} \label{AST}
    T \,\hat{=}\, - \frac{1}{2 \pi r (r^2 + a^2)} \left(\rho^4 (\mathrm{Re}(\Psi_2) + 2 \Lambda) - a^2 \left(1 - \frac{2 r^2 \sin^2\theta}{\rho^2}\right) \right),
\end{equation}
where we have considered the Penrose--Rindler equation \eqref{PR} to eliminate $\Phi_{11}$ in favor of $\mathrm{Re}(\Psi_2)$, $\Lambda$, and the Gaussian curvature \eqref{ASKG}, thus resulting in a simpler expression. Note that the temperature for static black holes given in Eq.~\eqref{SST} is recovered in the nonrotating limit $a \to 0$. We stress that the temperature \eqref{AST} is, of course, not dependent on the angular coordinate. The above arguments are valid as well. In fact, for simplification, the Kerr solution can be considered, and explicit expressions of the NP scalars can be substituted to prove that the result is the well-known Kerr temperature, which is $\theta$-independent.

We can finally repeat the same analysis of the interpretation of the Penrose--Rindler equation as a balance of pressures defined at the horizon, as done for spherically symmetric black holes previously, but now for axisymmetric black holes. Note that, in both cases, the Penrose--Rindler equation has been demonstrated to be equivalent to the horizon condition, which is indeed a remarkable result. We then write the Penrose--Rindler equation as in Eq.~\eqref{PR2}, namely,
\begin{equation} \label{PR3}
    - (\mathrm{Re}(\Psi_2) + 2 \Lambda) + \Phi_{11} + 3 \Lambda \,\hat{=}\, \frac{k_g}{2}.
\end{equation}
We can directly identify once again the pressure $P = - T_r^r$ associated with the radial component of the energy-momentum tensor, the matter pressure, as in Eq.~\eqref{P}, and the pressure $P_{k_g}$ associated with the Gaussian curvature, the curvature pressure, as in Eq.~\eqref{PKG}. However, $\mathrm{Re}(\Psi_2) + 2 \Lambda$ cannot be associated only with the thermal pressure as in Eq.~\eqref{PT} since the temperature \eqref{AST} now exhibits some extra terms. We have, in particular, that
\begin{equation}
    \frac{r (r^2 + a^2)}{2 \rho^4} T - \frac{a^2}{4 \pi \rho^4} \left(1 - \frac{2 r^2 \sin^2\theta}{\rho^2}\right) \,\hat{=}\, - \frac{1}{4 \pi} (\mathrm{Re}(\Psi_2) + 2 \Lambda).
\end{equation}
The above expression has two contributions: a contribution proportional to $T$ and an extra contribution proportional to $a^2$. We consider the contribution proportional to $T$ to define the pressure associated with the temperature, namely, the thermal pressure $P_T$, and the contribution proportional to $a^2$ to define the pressure associated with the rotation, namely, the rotation pressure $P_a$, hence,
\begin{equation} \label{PT+Pa}
    P_T + P_a \,\hat{=}\, - \frac{1}{4 \pi} (\mathrm{Re}(\Psi_2) + 2 \Lambda).
\end{equation}
Obviously, $P_a \to 0$ in the limit $a \to 0$, thus recovering Eq.~\eqref{PT} for spherically symmetric and static black holes. We can then conclude that stationary black holes with axial symmetry can be physically understood as defined by an equilibrium of horizon pressures given by
\begin{equation} \label{ASPEq}
    P \,\hat{=}\, P_T + P_a + P_{k_g}.
\end{equation}
Notice the presence of the new pressure term accounting for the black hole rotation, in contrast with the corresponding equilibrium for the nonrotating counterpart.

We ask, once that black hole horizon condition was shown to correspond geometrically to the Penrose--Rindler equation evaluated at the horizon, which physically reads like a horizon equilibrium of pressures now including a rotation term, whether a Smarr-like formula for arbitrary stationary black holes extending that for static black holes is recovered. In that case, the Penrose--Rindler equation \eqref{PR2} leads to the (generalized) Smarr formula \eqref{SSS} by multiplying it by $3 V$, where $V$ is the black hole geometric or thermodynamic volume, namely, the volume of the two-sphere. Let us now define the following quantity:
\begin{equation} \label{not_a_volume}
    V_\theta := \frac{4 \pi}{3} \frac{\rho^4}{r}.
\end{equation}
This quantity has dimensions of volume. However, $V_\theta$ is not a proper volume since the angular coordinate $\theta$ is present through $\rho^2 = r^2 + a^2 \cos^2 \theta$. Notice that, in the limit $a \to 0$, $V_\theta$ reduces to the volume of the two-sphere.

We consider now the Penrose--Rindler equation for axisymmetric black holes. One can check by multiplying Eq.~\eqref{PR3} by $3 V_\theta$ that $3 P_T V_\theta \,\hat{=}\, 2 T S$, but contrary to what one might naively expect, $3 P_{k_g} V_\theta \,\hat{\neq}\, - E$ and $3 P_a V_\theta \,\hat{\neq}\, 2 \Omega J$. As before, 
\begin{equation}
    E \,\hat{=}\, \frac{r^2 + a^2}{2 r}, \quad S = \frac{A}{4} \,\hat{=}\, \pi (r^2 + a^2)
\end{equation}
are the energy (generalization of the Misner--Sharp energy) and entropy of the horizon, respectively, and now
\begin{equation}
    J \,\hat{=}\, E a, \quad \Omega \,\hat{=}\, \frac{a}{r^2 + a^2}
\end{equation}
are the angular momentum and the angular velocity of the horizon, respectively \cite{Hansen:2016wdg}. The curvature pressure $P_{k_g}$ is then not directly related to the energy of the horizon in the presence of rotation, and, perhaps surprisingly, either the rotation pressure $P_a$ is exclusively tied in with the angular velocity of the horizon. Although this is true, it turns out that
\begin{equation}
     3 (P_a + P_{k_g}) V_\theta \,\hat{=}\, - E + 2 \Omega J,
\end{equation}
hence,
\begin{equation} \label{ASS}
    E \,\hat{=}\, 2 T S + 2 \Omega J - 3 P V_\theta,
\end{equation}
which generalizes the famous Smarr formula \cite{Smarr:1972kt} to arbitrary rotating black holes. Note that $P = - T_r^r$ is precisely the matter pressure, namely, the pressure associated with the radial component of the energy-momentum tensor (including a possible cosmological constant). We can deduce from Eq.~\eqref{ASS} that $V_\theta$ is the conjugate variable to $P = - T_r^r$, though $V_\theta$ is not interpreted as a black hole volume. As expected, the (generalized) Smarr formula for nonrotating black holes \eqref{SSS} is recovered when $a \to 0$ in Eq.~\eqref{ASS}.

Apart from the connection between the Penrose--Rindler equation and the Smarr formula considered before, the Penrose--Rindler equation interpreted as an equilibrium of pressures reproduced a van der Waals-like equation of state for static black holes once the holographic degrees of freedom of the horizon were introduced. In particular, the van der Waals equation of state was associated with the matter pressure, with the thermal pressure giving the ideal gas term and the curvature pressure giving the interaction term [see Eq.~\eqref{EoS}]. Notice that a black hole volume, conjugate to $P = - T_r^r$ in the Smarr formula \eqref{SSS}, was crucial for such an equation of state to show up. We have seen that an analogous notion of black hole volume does not exist in the presence of rotation, since the conjugate to $P = - T_r^r$ in the Smarr formula \eqref{ASS} is not a volume.

One could arbitrarily define an effective matter pressure $\tilde{P} = P \frac{V_\theta}{\tilde{V}}$ by introducing an effective volume $\tilde{V}$ so that the Smarr formula \eqref{ASS} reads
\begin{equation}
    E \,\hat{=}\, 2 T S + 2 \Omega J - 3 \tilde{P} \tilde{V},
\end{equation}
such that $\tilde{P}$ and $\tilde{V}$ are the conjugate variables, with $\tilde{V}$ a properly defined volume. The analysis of an effective equation of state of the form $\tilde{P} = \tilde{P}(\tilde{n}, T)$, with $\tilde{n} = \bar{N} / \tilde{V}$ and $\bar{N} = N / 6$, would be now possible. This study can be conducted through two alternatives approaches. One could identify the effective volume with a concrete volume for rotating black holes, for example, invoking the geometric or thermodynamic volume. Afterward, the analysis of the resulting equation of state can be done with the by-hand-introduced volume. This is the approach followed in \cite{Hansen:2016wdg}, but no microscopic considerations in terms of the horizon degrees of freedom were given that work. Another approach could be deriving the effective black hole volume such that the corresponding effective pressure satisfies the desired equation of state, for instance, the same van der Waals-like equation as for nonrotating black holes. Although effective, perhaps such equations of state are susceptible to emerge from an ensemble of spacetime atoms representing the holographic degrees of freedom of a rotating black hole. This was also the philosophy behind the van der Waals-like equation of state proposed for the Kerr--Newman solution specifically in \cite{Villalba:2020acp}.

The above alternatives are interesting but seem somehow naive and do not follow the spirit of horizon thermodynamics. We revisit this in the following, motivated by the new perspective provided by the GHP formalism.

\section{The Penrose--Rindler equation and the GHP formalism} \label{Sec4}

Let us move now to the GHP formalism \cite{Newman:1961qr, Geroch:1973am, Penrose1984, Chandrasekhar1998, Bargueno2023} in order to refine the Smarr-like relation \eqref{ASS}, where the ``volume'' term $V_\theta$ is angular dependent and has no immediate quasilocal interpretation. The GHP formalism extends the NP formalism by making it manifestly covariant under the so-called spin-boost transformations. In this case, the GHP formalism will inspire the introduction of a pressure-volume pair of conjugate thermodynamic variables entering the Smarr formula.

The GHP formalism preserves the structure of the NP null tetrad by considering the two real null vectors $\mathbf{l}$ and $\mathbf{n}$, understood as representatives of the equivalence classes $[\mathbf{l}] = \left\{\lambda \, \mathbf{l}\right\}$ and $[\mathbf{n}] = \left\{\lambda^{-1} \, \mathbf{n}\right\}$, respectively, with $\lambda \in C^{\infty}(M, \mathbb{R} \backslash \{0\})$, and the two complex conjugate null vectors $\mathbf{m}$ and $\overline{\mathbf{m}}$ taken to be tangential to the marginally trapped surfaces. The NP formalism is refined by restricting the class of allowable tetrad transformations to those preserving the null directions, which form the GHP group of spin-boost transformations \cite{Bargueno2023}. The GHP group is generated by the boosts
\begin{equation}
    \mathbf{l} \to \lambda \, \mathbf{l}, \quad \mathbf{n} \to \lambda^{-1} \, \mathbf{n},
\end{equation}
and the spatial rotations
\begin{equation}
    \mathbf{m} \to e^{i\theta} \, \mathbf{m},
\end{equation}
with $\theta \in C^\infty(M, \mathbb{R})$. A scalar $\eta$ is said to be a GHP scalar if it transforms covariantly under this Lorentz subgroup, where the transformation is $\eta \mapsto \lambda^p e^{i q \theta} \eta$, with $(p, q)$ its GHP weights. In this context, the NP derivatives along the null tetrad vectors are not covariant under the GHP group. As a consequence, the GHP formalism defines weighted derivatives that enlarge the NP derivatives to make them manifestly covariant under spin-boost transformations. The GHP derivative operators are $\thorn$ (GHP covariant derivative along $\mathbf{l}$) and $\eth$ (GHP covariant derivative along $\mathbf{m}$), as well as their primed ($'$) version, where the prime ($'$) indicates the interchange $\mathbf{l} \leftrightarrow \mathbf{n}$ and $\mathbf{m} \leftrightarrow \overline{\mathbf{m}}$.

The connection between Gaussian curvature and the Smarr formula has been recently explored for static and spherically symmetric spacetimes within the GHP formalism \cite{Guilabert:2024tga}. It was argued that the Smarr formula can be derived by integrating the Gaussian curvature over a cross-section of the trapping horizon, \textit{i.e.}, a future marginally outer trapped surface, and multiplying the result by a unit of energy. It turns out that, even before integration, the Gaussian curvature defines a natural notion of pressure in geometrized units. As already discussed, this pressure is related to the matter pressure introduced in Eq.~\eqref{P} via the Penrose--Rindler equation \cite{Penrose1984}:
\begin{equation} \label{PR4}
    \mathrm{Re}(\Psi_2) + 2 \Lambda + \mathrm{Re}(\rho \rho') - \mathrm{Re}(\sigma \sigma') + \frac{k_g}{2} = \Phi_{11} + 3 \Lambda,
\end{equation}
where (the real part of) the spin coefficient $\rho = m^\mu \delta' l_\mu$ encodes the expansion, and the spin coefficient $\sigma = m^\mu \delta l_\mu$ encodes the shear of the null congruence generated by $\mathbf{l}$ (outgoing null geodesics), with $\delta$ the NP derivative along $\mathbf{m}$. Notice that the same symbol $\rho$ is used both for the shorthand $\rho^2 = r^2 + a^2 \cos^2\theta$ and for the spin coefficient. The distinction should be clear from the context.

Let us clarify a couple of points here: (i) In the considered static and stationary spacetimes, the null congruence generated by $\mathbf{l}$ is such that $\sigma = 0$. (ii) At the black hole horizon, the null congruence generated by $\mathbf{l}$ is such that $\rho = 0$. The black hole trapping horizon is a null surface, and its foliation by future marginally outer trapped surfaces verifies $\rho = 0$. The Penrose--Rindler equation \eqref{PR3} considered so far therefore follows from Eq.~\eqref{PR4} and these two points, leading to the geometric decomposition of the matter pressure \eqref{P} given by
\begin{equation} \label{PEq}
    P \,\hat{=}\, - \frac{1}{4 \pi} (\mathrm{Re}(\Psi_2) + 2 \Lambda) - \frac{k_g}{8 \pi},
\end{equation}
which suggests the introduction of the curvature pressure. As discussed before, $\mathrm{Re}(\Psi_2) + 2 \Lambda$ cannot be, in general, considered as a contribution related to the black hole temperature. This is only possible in the static and spherically symmetric case [see Eq.~\eqref{PT}]. In the stationary and axisymmetric case, $\mathrm{Re}(\Psi_2) + 2 \Lambda$ encompasses the thermal contribution together with extra geometric contributions related to rotation [see Eq.~\eqref{PT+Pa}].

The connection between geometry and temperature turns out to arise from the behavior of the real part of the spin coefficient $\rho$. As argued by Hayward, the black hole temperature is linked to the real part of the rate of change of $\rho$ along the ingoing null direction, more specifically, to $\mathrm{Re}(\thorn' \rho)$ \cite{Hayward:1994yy} (see also \cite{Fodor:1996rf, Nielsen:2007ac, Guilabert:2024tga}). In the GHP formalism, this is encoded in the equation
\begin{equation} \label{GHP}
    \thorn' \rho - \eth' \tau = \rho \overline{\rho}' + \sigma \sigma' - \tau \overline{\tau} - \kappa \kappa' - (\Psi_2 + 2 \Lambda),
\end{equation}
where the spin coefficient $\tau = m^\mu D' l_\mu$ encodes the extrinsic rotation relative to transverse surfaces, which is physically related to frame dragging, and the spin coefficient $\kappa = m^\mu D l_\mu$ encodes the transverse (nongeodesic) acceleration of the null congruence generated by $\mathbf{l}$ (outgoing null geodesics), with $D$ the NP derivative along $\mathbf{l}$.

In the case of the static and spherically symmetric black holes described by the metric \eqref{SSMetric}, Eq.~\eqref{GHP} considerably simplifies to
\begin{equation}
   \thorn' \rho \,\hat{=}\, - (\Psi_2 + 2 \Lambda),
\end{equation}
since the other terms vanish. As a consequence, Eq.~\eqref{PEq} gives
\begin{equation}
    P \,\hat{=}\, \frac{1}{4 \pi} \thorn' \rho - \frac{k_g}{8 \pi},
\end{equation}
which is just Eq.~\eqref{SSPEq}.

In the case of the stationary and axisymmetric rotating (Kerr-like) black holes described by the metric \eqref{ASMetric}, the real part of Eq.~\eqref{GHP} reduces to
\begin{equation}
    \mathrm{Re}(\thorn' \rho) - \mathrm{Re}(\eth' \tau) + \tau \overline{\tau} \,\hat{=}\, - (\mathrm{Re}(\Psi_2) + 2\Lambda),
\end{equation}
so Eq.~\eqref{PEq} yields
\begin{equation}
    P \,\hat{=}\, \frac{1}{4 \pi} \mathrm{Re}(\thorn' \rho) - \frac{1}{4 \pi} \mathrm{Re}(\eth' \tau) + \frac{1}{4 \pi} \tau \overline{\tau} - \frac{k_g}{8 \pi}.
\end{equation}
The above expression clearly differentiates four contributions, so suggesting a further geometric decomposition of the pressure including terms associated with the Gaussian curvature, the temperature, and two additional contributions linked to the rotation of the null directions. We have that
\begin{equation} \label{GHPPEq}
    P \,\hat{=}\, P_T + P_{\eth' \tau} + P_{\tau \overline{\tau}} + P_{k_g},
\end{equation}
where $P_T \,\hat{=}\, \frac{1}{4 \pi} \mathrm{Re}(\thorn' \rho)$ and with the rotation terms divided into $P_{\eth' \tau} = - \frac{1}{4 \pi} \mathrm{Re}(\eth' \tau)$ and $P_{\tau \overline{\tau}} = \frac{1}{4 \pi} \tau \overline{\tau}$. The comparison of Eq.~\eqref{ASPEq} with Eq.~\eqref{GHPPEq} demonstrates that the rotation pressure $P_a$ previously introduced is such that $P_a \,\hat{=}\, P_{\eth' \tau} + P_{\tau \overline{\tau}}$. The GHP formalism then allows a more precise decomposition of the pressure in purely geometric terms with transparent interpretation, thus justifying its use.

Notably, among the terms in Eq.~\eqref{GHPPEq}, the only one that depends explicitly on the details of the geometry, specifically on $\Delta'(r)$, is the thermal pressure $P_T$. As a consequence, for two Kerr-like black holes possibly arising from different gravity theories, the nonthermal pressure terms are the same. This fact indicates that the radial Einstein field equation at the horizon involves only a single ``degree of freedom'' given by the requirement $\Delta(r) \,\hat{=}\, 0$.

One can show that the thermal pressure $P_T$ indeed is
\begin{equation}
    P_T \,\hat{=}\, \frac{1}{4 \pi} \mathrm{Re}(\thorn' \rho) \,\hat{=}\, \frac{r (r^2 + a^2)}{2 \rho^4} T.
\end{equation}
As already realized,
\begin{equation}
    P_T \,\hat{=}\, \frac{2 T S}{3 V_\theta},
\end{equation}
in terms of the black hole entropy and $V_\theta$ \eqref{not_a_volume}.

We recall that Eq.~\eqref{GHPPEq} has no angular dependence as a whole. We choose here, in order to remove the $\theta$-dependence in each individual pressure term, to integrate it over the horizon and then divide by the area of the horizon. Note that $P = - T_r^r$ and $T_r^r$ depends on $\theta$, so the introduction of an average of the radial component of the energy-momentum tensor at horizon is well motivated. This averaging procedure is not just convenient for our purposes. It is common in the framework of quasilocal geometry \cite{Hawking:1968qt, Penrose:1982wp, Brown:1992br, Wang:2008jy, Szabados:2009eka}. It is also consistent with black hole thermodynamics, where quantities such as entropy and temperature can be defined through integrals over marginally trapped surfaces \cite{Wald:1993nt, Ashtekar:2004cn, Guilabert:2024tga}. Averaging $P$ over the horizon will then produce a quantity that can be compared across slicings. This average will therefore promote $P$ to a quasilocal quantity intrinsically related to the horizon. What is more, remarkably, the pressure term associated with $\eth' \tau$ is identically zero when integrated over a closed horizon section \cite{Penrose1984, Hayward:1994yy}, making this a physically meaningful prescription.

We denote the average value of the pressure $P$ as $\langle P \rangle$, so that
\begin{equation}
    \langle P \rangle := \frac{1}{A} \int_H \star P,
\end{equation}
where $\star$ denotes the Hodge dual on the marginally trapped surface $H$, with the area of the black hole horizon given by
\begin{equation}
    A = \int_H \star 1.
\end{equation}
Note that, by doing this, we will be describing the thermodynamic system effectively. As a horizon-averaged description, any anisotropies or local variations in the distribution of the pressure over the horizon are lost in $\langle P \rangle$, making the effective thermodynamic system not equivalent to the initial one. 

As already mentioned, the average pressure $\langle P_{\eth' \tau} \rangle$ vanishes, since $\eth' \tau$ is a total derivative integrated over a compact surface without boundary \cite{Penrose1984, Hayward:1994yy}. As a consequence, the averaged effective pressure decomposition simplifies to
\begin{equation} \label{GHPPEq2}
    \langle P \rangle \,\hat{=}\, \langle P_T \rangle + \langle P_{\tau \overline{\tau}} \rangle + \langle P_{k_g} \rangle,
\end{equation}
with
\begin{equation}
    \langle P_T \rangle \,\hat{=}\, \frac{2 T S}{3} \left\langle \frac{1}{V_\theta} \right\rangle.
\end{equation}
Let us define the effective volume $V$ at this stage, such that the inverse of $V$ is the average of the inverse of $V_\theta$, namely,
\begin{equation}
    \frac{1}{V} := \left\langle \frac{1}{V_{\theta}} \right\rangle,
\end{equation}
obtaining
\begin{equation}
    V = \frac{4 \pi r^3}{3} \frac{2 \left(r^2 + a^2\right)}{r^2 + \frac{r}{a} \left(r^2 + a^2\right) \tan^{-1}(\frac{a}{r})},
\end{equation}
recovering the volume of the two-sphere in the limit $a \to 0$. This allows us to rewrite Eq.~\eqref{GHPPEq2} in the form
\begin{equation}
    3 \langle P \rangle V \,\hat{=}\, 2 T S + 3 (\langle P_{\tau \overline{\tau}} \rangle + \langle P_{k_g} \rangle) V.
\end{equation}
Analogous to the previous section, $3 \langle P_{k_g} \rangle V \,\hat{\neq}\, - E$ and $3 \langle P_{\tau \overline{\tau}} \rangle V \,\hat{\neq}\, 2 \Omega J$, but
\begin{equation}
     3 (\langle P_{\tau \overline{\tau}} \rangle + \langle P_{k_g} \rangle) V \,\hat{=}\, - E + 2 \Omega J,
\end{equation}
hence,
\begin{equation} \label{S}
    E \,\hat{=}\, 2 T S + 2 \Omega J - 3 \langle P \rangle V,
\end{equation}
thus recovering the Smarr formula for Kerr-like black holes. Note that this gives $V$ a clear thermodynamic interpretation as the conjugate variable to the averaged radial pressure over the horizon. We will refer to this volume as the Smarr volume. It is also clear that the Smarr volume differs from the geometric and thermodynamic volumes for Kerr-like black holes \cite{Kubiznak:2016qmn}. In particular, the latter do not satisfy the Smarr formula \eqref{S}. As argued before, one can force them to verify an equation of the form $E \,\hat{=}\, 2 T S + 2 \Omega J - 3 \tilde{P} \tilde{V}$, but the conjugate effective pressure $\tilde{P}$ has no energy-momentum tensor transparent interpretation, losing somehow the spirit of horizon thermodynamics. The Smarr volume is a natural volume candidate that satisfies the Smarr formula \eqref{S} with a clear interpretation for the conjugate effective pressure $\langle P \rangle$ in the context horizon thermodynamics as the average of the radial component of the energy-momentum tensor.

The Smarr volume can also be understood as the consequence of the quasilocal geometry in the effective thermodynamic description of the black hole due to the horizon rotation, however, improving the presentation of Kerr-like black holes thermodynamically in pressure-volume terms. It allows, in this sense, the definition of a quasilocal equation of state for the horizon degrees of freedom $\langle P \rangle = \langle P \rangle (n, T)$, which would be consistent with the Smarr formula \eqref{S} and with the philosophy behind the horizon approach to rotating black hole thermodynamics. The analysis of this equation of state is under development and will be presented elsewhere.

\section{Final remarks} \label{Sec5}

This manuscript presents a coherent framework for geometrically understanding black hole thermodynamics using the NP and the GHP formalisms. We have shown that the black hole condition is equivalent to the Penrose--Rindler equation at the horizon, which itself provides an alternative formulation the radial Einstein field equation evaluated at the horizon. This allows a physical understanding of black holes as a horizon equilibrium of pressures with transparent geometric interpretation in terms of local geometric quantities. Additionally, the decomposition into curvature, thermal, and rotation pressures supporting such an equilibrium reveals that only the temperature-dependent term depends explicitly on the detailed radial spacetime structure. The Penrose--Rindler equation was furthermore shown to be equivalent to the generalized Smarr formula, the master equation of black hole thermodynamics, thus highlighting the interplay between geometry and thermodynamics. 

This approach extends and unifies horizon thermodynamics and, moreover, motivates a quasilocal realization of the Smarr relation for stationary and axisymmetric rotating (Kerr-like) black holes within this context. This is based on the introduction of the Smarr volume, a horizon-averaged candidate, conjugate to the averaged radial component of the energy-momentum tensor, which plays the role of the matter pressure in horizon thermodynamics. Unlike the typically considered geometric or thermodynamic volumes, the Smarr volume is constructed to define the quasilocal Smarr formula \eqref{S}, hence encapsulating the effective thermodynamic response of the rotating horizon. As expected, everything reduces consistently to the well-known results for static spacetimes in the limit of no rotation.

This approach permits several natural extensions. First of all, our analysis is amenable to be extended to Kerr--(anti-)de Sitter-like spacetimes straightforwardly. In this case and in the context of horizon thermodynamics, the cosmological constant defines an additional contribution to the matter sector, hence entering the quasilocal pressure here introduced. The same Smarr-like will follow with a Smarr volume that will now also be a function of the cosmological constant. Lastly, our analysis applies similarly to modified gravity theories such as $f(R)$ gravity or Lovelock theories, since the NP and GHP formalisms are independent of the specific gravitational theory.

Finally, an interesting investigation within this approach is that of the corresponding quasilocal equation of state for Kerr-like black holes, following from the generalized Smarr formula \eqref{S}, in terms of the holographic degrees of freedom of the horizon. We defer the details to future work.

\section*{Acknowledgments}

DFS appreciates valuable discussions with Martina Adamo. DFS is partially supported by the Grant No. PID2023-148373NB-I00 funded by MCIN/AEI/10.13039/501100011033/FEDER, UE, by the Q-CAYLE Project funded by the Regional Government of Castilla y León (Junta de Castilla y León), and by the Ministry of Science and Innovation (MCIN) through the European Union funds NextGenerationEU (PRTR C17.I1). PB and JAM acknowledge financial support from the Generalitat Valenciana through PROMETEO PROJECT CIPROM/2022/13.

\bibliography{main}

\end{document}